\documentclass[12pt]{article}
\usepackage{graphicx}
\usepackage{color}
\newcommand{\beq}{\begin{equation}}
\newcommand{\eeq}{\end{equation}}
\newcommand{\bea}{\begin{eqnarray}}
\newcommand{\eea}{\end{eqnarray}}

\newcommand{\gsim}{\lower.7ex\hbox{$
\;\stackrel{\textstyle>}{\sim}\;$}}
\newcommand{\lsim}{\lower.7ex\hbox{$
\;\stackrel{\textstyle<}{\sim}\;$}}

\newcommand{\eod}{\end{document}}

\begin{document}
\thispagestyle{empty}
\vspace*{-22mm}

\begin{flushright}
UND-HEP-15-BIG\hspace*{.08em}05\\
Version 6
 \\


\end{flushright}
\vspace*{1.3mm}

\begin{center}
{\Large {\bf Is There a Real Difference between $|V_{ub}|_{\rm excl}$ vs. $|V_{ub}|_{\rm incl}$ 
from Semi-leptonic $B$ Decays?}}

\vspace*{10mm}

{Ikaros I.~Bigi}\\
\vspace{5mm}
{\sl Department of Physics, University of Notre Dame du Lac, Notre Dame, IN 46556, USA}\\

{\sl E-mail address: ibigi@nd.edu} \\

\vspace*{5mm}

{\bf Abstract}
\vspace*{-1.5mm}
\\
\end{center}

For a long time there have been active discussions  about the sizable differences 
between $|V_{ub}|_{\rm incl}$ \& $|V_{ub}|_{\rm excl}$. However, the real $|V_{ub}|_{\rm incl}$ 
might be smaller than usually stated.   
The connection of the worlds of 
hadrons and quarks (\& gluons) is subtle: we cannot ignore final states with a pair 
of $\bar KK$ mesons plus pions: 
$\left[ R (B \to l \nu \; \pi ' s)  +  R (B \to l \nu \; (K\bar K +\pi 's ))  \right]$ 
$\simeq$  $R ([b\bar q] \to l \nu \, u (\bar q_i q_i + \bar q_jq_j\bar q_k q_k...) \bar q )$ with $\bar q=\bar u, \bar d$ \&  
$q_i = u,d,s$. 
It might show the limits or violations of duality close to thresholds. 
While inclusive FS cannot be measured, exclusive ones can be done  
soon by LHCb and later also by Belle II about $B^+ \to l^+ \nu K^+K^-$ \&  
$B^0 \to l^+ \nu K^+K^-\pi^+$. Present data have not given us the information 
about resonances in the region of 1 - 2 GeV that we need to understand the underlying dynamics 
in $\Delta S=0$. 
Future data for exclusive $B \to l \nu \pi \pi$ might also narrow the gap between 
$|V_{ub}|_{\rm incl}$ \& $|V_{ub}|_{\rm excl}$ in the opposite direction.   
I comment about $B_s^0 \to l^+ \nu X_u^{(s)}(\Delta S = -1)$.

\vspace{3mm}

\hrule

\tableofcontents
\vspace{5mm}

\hrule\vspace{5mm}

\section{The `Problem'}
\label{INTRO}

There is a `long' history about the differences between exclusive vs. inclusive semi-leptonic 
decays of $B$ mesons. I focus on $B \to l \nu \pi/\rho$ vs. $B\to l \nu X_u$ leading to 
$|V_{ub}|_{\rm excl}$ $<$ $|V_{ub}|_{\rm incl}$, but only mention   
$B\to l \nu D^{(*)}$ vs. $B \to l \nu X_c$.  
Literature use different tools, which is enough for this exercise; for more details 
one can find it in PDG2014:  
\bea
|V_{ub}|_{\rm incl}   & \simeq & (4.41\pm 0.15 ^{+0.15}_{-0.17}) \cdot 10^{-3} \\    
|V_{ub}|_{\rm excl}   & \simeq & (3.28 \pm 0.29) \cdot 10^{-3} \; .
\eea
It is important to see the details how one needs to produce the numbers \cite{PhBFacts}:
\bea
|V_{ub}|_{\rm incl}   & \simeq & 4.42(1\pm 0.045_{\rm exp} \pm 0.034_{\rm th})  \cdot 10^{-3} \\    
|V_{ub}|_{\rm excl}   & \simeq & 3.23(1 \pm 0.05_{\rm exp} \pm 0.08_{\rm th}) \cdot 10^{-3} \; .
\eea
It is not clear why one should average these two values. We realize that the situation is 
complex, as you can see in the literatures:  
\bea    
|V_{ub}|_{\rm excl}   & \simeq & (3.7 \pm 0.2) \cdot 10^{-3} \; \; {\rm MILC} \\
|V_{ub}|_{\rm excl}   & \simeq & (3.3 \pm 0.3) \cdot 10^{-3} \; \; {\rm HFAG}
\eea
MILC describes form factor of $\langle \pi |J_{\mu}|B\rangle$ from LQCD, while HFAG uses  
exclusive data. Furthermore one can probe the `golden triangle' based on correlations leading to:
\beq
|V_{ub}| \simeq (3.45^{+0.23}_{-0.10}) \cdot 10^{-3} \; \; {\rm CKMfitter}
\eeq 
Some will say these numbers show sizable differences between the values from one `camp' on the higher side 
of 0.004 and those from the three `camps' on the lower one. While the central numbers have hardly changed in ten years, 
the gaps have enhanced.
Several authors of Refs.\cite{PhBFacts,MANNEL,MEMORIAL} 
have thought and worked on both sides of this `Problem'. 

Numbers are not always give the best information we can get from the data. 
In my view there are three camps to describe final states (FS) with somewhat different tools: 
inclusive ones from $b \to u l^- \bar \nu $; 
MILC and HFAG describe similar landscapes with different tools; 
results from CKMfitter are based on correlations with other transitions.

We can learn from this situation with three options: 
(a) We do not understand how to deal with inclusive transitions. 
(b) It has been suggested that we might have underestimated the uncertainties in $B \to l \nu \pi$. One was 
discussed in Ref.\cite{MANNEL} about disentangling observables in $B^- \to l^- \bar \nu \pi^+\pi^-$. It is not trivial 
to measure them, but using tools like HQET, SCET etc. give us more information, if they are within their limits. 
It was discussed also to probe $B^- \to l^- \bar \nu \pi^+\pi^-$ based on dispersion relations 
to get model-insensitive analyses \cite{KANG}. 
(c) ND is hiding in the correlations \& its uncertainties of CKMfitter. 

Personally I find options (c) \& (b) very interesting. Maybe I am biased; 
however, I see no reason, why option (a) is a real one. 

I add a comment about a very interesting paper \cite{STEFAN}. They are courageous to solve 
both the differences between $|V_{cb}|_{\rm excl}$ vs. $|V_{cb}|_{\rm incl}$ and $|V_{ub}|_{\rm excl}$ vs. 
$|V_{ub}|_{\rm incl}$.  
They found in the `global' landscape they considered there is hardly any chance to find ND without too 
much impact on the measured widths of $Z \to \bar bb$. 

Instead I focus on $|V_{ub}|_{\rm excl}$ vs. $|V_{ub}|_{\rm incl}$ on a narrow landscape; we have to measure 
semi-leptonic decays of $B$ mesons with two \& three light hadrons in the final states (FS). 
My main point is that "duality" is more subtle. Thus I do not give up on option (c) yet. 

We can use heavy quark parameters extracted from $B \to l \nu X_c$. It is not trivial, 
but we know how to do it. 
There is another challenge, namely to compare $B^+ \to l^+  \nu X^0_u$ vs. 
$B^0 \to l^+  \nu X^-_u$ \cite{74}. It is not easy, but there is a prize: probing the impact 
of `weak annihilation' (WA) give us novel lessons about non-perturbative QCD (and more). 
Furthermore $\Gamma(B \to l \nu X_u)$ are under better theoretical control than 
$\Gamma(B \to l \nu X_c)$ in the SM, since the expansion parameter is smaller  
($\frac{\mu}{m_b}$ vs. $\frac{\mu}{m_b - m_c}$), and ${\cal O}(\alpha^2_S)$ corrections are known.  
In the real world there arise more challenges: to distinguish $b\to u$ from huge $b\to c$ backgrounds 
one applies cuts on variables like charged lepton energy $E_l$, hadronic mass $M_X$ and the lepton-pair 
invariance mass $q^2$ to `manageable' proportions.  
We have learnt from detailed analyses of radiative $B \to \gamma X_s$ decays \cite{94}.
Furthermore one has  to deal with other 
backgrounds that are not connected with the production of beauty hadrons. 

BaBar and Belle have measured the energy of the charged lepton 
in the region of $\sim$ 2.0 - 2.6 GeV. When one gets close to $\sim$ 2.0 GeV from 
{\em above}, it is more difficult to measure that. 
Below one has to depend on the extrapolation {\em down} from the measured region; one expects a
long tail of $|V_{ub}|$ amplitudes. The questions are: how much and where. 
The state-of-the-art theoretical tools had been applied to $M_X$ \& $q^2$ in $B \to l \nu X_u$ \&  
also in $B \to \gamma X_s$ and discussed in details \cite{PAOLO}.  
Measuring the hadronic recoil spectrum up to a maximal values $M_X^{\rm max}$ captures the lion 
share of the $b \to u$ rate if $M_X^{\rm max}$ is below 1.7 GeV; yet it is still vulnerable  
to theoretical uncertainties in the low-$q^2$ region. One can think to apply combinations, 
and it has been done for $|V_{ub}|_{\rm incl}$ in the world of quarks \cite{PAOLO}. 

It has been suggested to measure semi-leptonic $B \to l \nu \pi \pi$
\cite{KANG} to test the number of $|V_{ub}|_{\rm excl}$ from $B \to l \nu \pi$ using dispersion relations. This class of  
tools has a long history to describe three-body FS with strong \& electromagnetic forces (including chiral symmetry) like $\eta \to 3 \pi$ or 
$\eta ^{(\prime)} \to \pi \pi \gamma$.

The usually stated value of $|V_{ub}|_{\rm incl}$ is based on "duality" in HEP. 
In this paper I suggest that its  real value might be smaller and thus narrows the gap between $|V_{ub}|_{\rm excl}$ and $|V_{ub}|_{\rm incl}$: 
in semi-leptonic decays of $B$ mesons might not enter the region where semi-local duality works with good accuracy. The reason? 
To be honest: good/bad luck due to threshold(s) in the region of 1 - 2 GeV with resonances with $\Delta S=0$ decaying to $\bar K K$ plus 
$\pi^{\prime}$s
produce unusually impact. I cannot predict that, but miracles can happen -- rarely. After working hard and with successes about 
semi-leptonic inclusive rates, I cannot give up easily; instead I suggest we have to follow the data in several ways, which I discuss below. 
There is another challenge, namely to discuss the different `cultures' in Hadrodynamics \& HEP and their definitions of the item "duality". 
Still fundamental dynamics are formulated in the world of quarks, not 
pions, kaons \& $\eta^{(\prime)}$ and their resonances; therefore "duality" means for 
connections between the worlds of quarks \& hadrons.

\section{Duality: subtle tools \& their limits}
\label{DUAL}

Operator Product Expansion (OPE) combined with heavy quark expansion (HQE) 
have been applied to quantify non-perturbative effects in a number of important processes since the early 1980's \cite{MISHA} including  
heavy flavor hadrons.  
These papers and others had mostly focused on inclusive $b \to c$ amplitudes for semi- \& non-leptonic 
decays of beauty hadrons with final states (FS) of $D$ and resonances $D^*$, $D^{**}$ etc.  
and the perturbative domain. Later OPE \& HQE had been applied also to $b\to u$ amplitudes with very refined tools like in Refs.\cite{PAOLO} 
and others. Often the impact of WA amplitudes was ignored, but not in Ref.\cite{PAOLO}. It might produce a difference in the 
endpoint of semi-leptonic $B^+$ vs. $B^0$ transitions; however present analyses show little impact of WA 
\cite{TWOROADS,PhBFacts}.

It was explained in Ref.\cite{VADEMECUM} in the beginning of this century why duality can{\em not} be seen as additional assumption 
beyond quantum field theory. 
The landscape of quark-hadron duality can be subtle: 
in the regions of thresholds one has to apply semi-averaging (or `smearing')  
duality in intelligent ways with a well-known example:  we describe 
$e^+e^- \to H_c \bar H_c^{\prime} +$ light flavor hadrons in the region of $\psi(3770)$: below 
we have two very narrow resonances $\psi(1S)$ \& $\psi(2S)$. 
Pairs of open charm mesons appear just below $\psi(3770)$.Thus there are two gaps 
in the description of hidden \& open charm hadrons: 
a small one between $\psi(2S)$ \& just below $\psi(3770)$; a large one between $\psi(1S)$ \& $\psi(2S)$. In the world of charm quarks with 
$m_c \sim 1.3$ GeV \footnote{ PDG2014 gives us for the charm quark $m_c = (1.275 \pm 0.025)$ GeV.}
without gaps in the world of quarks 
\footnote{As well known the definition of quark masses is subtle. 
We can{\em not} use pole masses as explained in details \cite{POLEQUARK}; instead we have to use scale depending value like for `kinetic' quark masses.}. 
It tells us that we have to use {\em semi}-averaged duality over 
a region $\sim$ 1 - 1.5 GeV in the world of quarks. Do we have enough space to describe well semi-leptonic decays of beauty mesons  
Still duality gives unitarity in subtle ways. 

In the world of quarks one uses amplitudes $T([b\bar q] \to l^- \bar \nu u\bar q)$ to describe inclusive 
transitions in $T(B \to l^- \bar \nu [\pi's + \bar K K \pi's])$. Of course, it is easy to add 
internal pairs of $u$, $d$ \& $s$ (anti-)quarks to connect with the world of hadrons.  
As first guess one might suggest ratios of $\bar u u /\bar d d \simeq 1$ 
and $\bar s s /\bar u u \sim 1/3 - 1/2$. It is naive to say that $\sim$ 14 - 20 \% of the FS 
comes from $(u\bar s)(s\bar q)$ and therefore for $K \bar K +$ pions. Anyway, it enhances the averaged mass of $X_u (\Delta S=0, -1)$. 
We have a better understanding of that due to mixing of 
$\langle 0|\bar u u|0\rangle$, 
$\langle 0|\bar d d |0\rangle$ between $\langle 0|\bar s s |0\rangle$ with scalar resonances that are not OZI suppressed \cite{DR}, 
but so far not quantitatively. Its impact could be smaller. 
However, the landscape is even much more complex.

Since I had worked with applying OPE \& HQE (and trust them), I might be seen as biased; 
however, I know that experienced HEP people have thought about them and discussed without solving the `Problem' \cite{PhBFacts}. 
It seems it could be a good time to look at unusual ways, although the probability to solve that is sizably less than 50 \% as I admit.

\section{Idea about a new road}
\label{NEWIDEA}

Decays of $B \to l^+ \nu + X_c$ give mostly described by a few hadrons in the FS, namely $D$, $D^*$ and narrow $D^{**}$ resonances. 
The situation is more `complex' in $B \to l^+\nu + X_u$ decays with many hadrons in the FS (and large background).    
I `paint' the landscape (when I refer to diagrams) to make the value of $|V_{ub}|_{\rm incl}$ smaller. Of course, the `primary intermediate' 
$u \bar q$ (with $\bar q = \bar u,\bar d$)
produce `secondary' $u \bar q_i q_i \bar q_j q_j...\bar q$ \& $q_{i,j}$ ones with $u,d,s$ 
before they give FS $K \bar K +\pi's$ with 
non-zero probability due to (strong) final states interactions (FSI) (or re-scattering).  
The question is what does that mean quantitatively? 

Adding a pair of $\bar ss$ to intermediate $u \bar q$ (or $u\bar s s \bar q_1q_1\bar q$ ...) is one thing, 
but understanding the dynamics is quite another thing. Resonances happen and have impact;  
one cannot describe that 
with local operators in general. Close to thresholds we have to deal with semi-averaged duality; 
its impacts are less predictable. 
They depend on the features of the FSI, 
not just their existence \footnote{Actually the definitions of limits or violations of duality are fuzzy.}. 

Data list in PDG2014 show BR$(B^{0,+} \to l^+ \nu X_u(\Delta S=0)) \sim 2 \cdot 10^{-3}$. However, 
they depend on extrapolations down 
from the measured region. Most people think that inclusive semi-leptonic decays of 
$B^{0,+}$ mesons consist basically of FS with only pions (+ $\eta^{(\prime}$). 
I talk about the combination of two items: 
(i) $\Delta S= 0$ resonances that produce a pair of strange mesons plus pions in the region of 1 - 2 GeV.  
(ii) Their impacts might be enhanced being close to thresholds. Those can go to 
$B^+ \to l^+ \nu [K^+K^-/K^+K_S\pi^-/K^-K_S\pi^+/K^+K^-\pi^+\pi^-/...]$ and 
$B^0 \to l^+\nu [K^+K^-\pi^-/K^-K_S\pi^+\pi^-/...]$. I focus on FS that LHCb collab. can probe.

\subsection{Impact of $K \bar K$, $K \bar K \pi$ etc. resonances}
\label{IMP}

Obviously LHCb can{\em not} measure inclusive transitions. Maybe also Belle II cannot solve the 
challenge of "$|V_{ub}|_{\rm incl}$" $ >|V_{ub}|_{\rm excl}$ directly. I suggest that limits or violation of 
duality  can be larger than expected due to combine two items: resonances produce  
sizable pairs of $\bar KK$ \& $\bar KK \pi$ can be enhanced by being 
close to thresholds. Maybe just luck? 
Re-scattering \cite{1988BOOK,WOLFFSI,CPBOOK} is crucial for 
$u \bar u \to s \bar s$, $s\bar q q \bar s$ etc., which are {\em not} described by local operators. 
It seems to me unlikely to measure {\em inclusive} decays $B \to l \nu K \bar K + \pi'$'s due a huge background from $B \to l \nu X_c$ and other sources.

\subsubsection{Exclusive semi-leptonic $B^{0,+}$ decays with $\Delta S=0$}

So far we have measured exclusive rates with a (pseudo-)single hadron in the FS including well-known resonances:  
\bea
{\rm BR} (B^+ \to l^+ \nu \omega ) = (1.19 \pm 0.09) \cdot 10^{-4} &,&
{\rm BR} (B^+ \to l^+ \nu \rho^0 ) = (1.58 \pm 0.11) \cdot 10^{-4} 
\nonumber \\
{\rm BR} (B^+ \to l^+ \nu \pi^0 ) &=& (7.80 \pm 0.27) \cdot 10^{-5} 
\nonumber \\
{\rm BR} (B^+ \to l^+ \nu \eta ) = (3.8 \pm 0.6) \cdot 10^{-5} &,&
{\rm BR} (B^+ \to l^+ \nu \eta^{\prime} ) = (2.3 \pm 0.6) \cdot 10^{-5} 
\eea
It was pointed out  in Ref.\cite{GIULIA} that a more definite conclusion has {\em not} been found yet 
about the impact of gluonia contributions to $\eta^{(\prime)}$ amplitudes. In my view that is not an 
academic discussion.  I will come back to that item below 
\footnote{ The landscape of $B^+ \to l^+ \nu p \bar p$, $B^0 \to l^+ \nu p \bar p$, 
$B^0 \to l^+\nu p \bar p K^0$ is very 
interesting about the impact of FSI \& non-perturbative QCD, but these data show no impact on 
measuring   $|V_{ub}|_{\rm incl}$.}.
\bea
{\rm BR} (B^0 \to l^+ \nu \pi^- ) &=& (1.45 \pm 0.05) \cdot 10^{-4}
\nonumber \\
{\rm BR} (B^0 \to l^+ \nu \rho^- ) &=& (2.94 \pm 0.21) \cdot 10^{-4} \; . 
\eea
Both $B^+$ \& $B^0$ produce around 20 \% of the inclusive ones in measured exclusive ones. 
On the other hand multi-body FS give around 80 \%   
including both narrow \& broad resonances.

Diagrams show with $\bar q = \bar u, \bar d$, added pairs of (anti-)$q_{i,j,k} = u,d,s$ in the FS 
and connect with hadrons:
\bea 
[b \bar q] \Rightarrow l \nu \, [u\bar q_i][ q_i \bar q]  &\simeq & B \to l \nu \, ( \pi \pi/\bar KK ) \cr
[ b \bar q ] \Rightarrow l \nu \, u\bar q_i q_i \bar q_j q_j \bar q  &\simeq&  B \to
l \nu \, (3 \pi/\bar K K \pi ) \cr 
[ b \bar q ] \Rightarrow l \nu \, u\bar q_i q_i \bar q_j q_j \bar q_k q_k \bar q  &\simeq&  B \to
l \nu \,  (4 \pi/\bar KK 2 \pi) 
\eea
Some of my comments are based on hand-waving arguments now, but not all. 

(a) In these quark diagrams I use "$\Rightarrow$" instead of the usual "$\to$" to emphasize the impact of 
FSI; in general those cannot be described with local operators. 

(b) I expect that 
LHCb can probe $B^+ \to l^+\nu K^+K^-$ and $B^0 \to l^+\nu K^+K^- \pi^- $ on the level of branching ratios of 
a few$\times 10^{-4}$. It might narrow the gap between $|V_{ub}|_{\rm incl}$ and $|V_{ub}|_{\rm excl}$ claimed before.  
If nothing was found there, it is the end of my idea. 
However, {\em if} LHCb will find {\em non}-zero data about $B^+ \to l^+\nu K^+K^-$ \& 
$B^0 \to l^+\nu K^+K^- \pi^-$, 
we have to think how much values we expect from limits or violations of duality. It is possible to predict $B\to l\nu K \bar K$ decays; however it would need 
large work analyzing such FS like based on dispersion relations (\& hoping to apply chiral symmetry). 
I do not suggest to work 
on a project, unless new data show the roads to understand such exclusive FS. 

(c) We cannot stop at $\pi$, $\rho$, $\omega$ \& $\eta^{(\prime)}$. 
We have to analyze $\pi \pi$ in general as pointed out \cite{MANNEL,KANG}; likewise to go beyond $\omega$. 
Measuring exclusive $B \to l^+ \nu [3 \pi, 4 \pi]$ would help our understanding of duality 
semi-quantitatively and its limits at least.

\subsubsection{Future data about resonances with $\Delta S=0$}

Actually there are still different roads leading to the goal. The PDG2014 shows that there are many resonances in the region of 1 - 2 GeV that 
could contribute in this landscape. Actually there are four classes now: 
\begin{itemize}
\item 
One basically produces only pions. 

\item
The second one gives some $\bar KK$ pairs like $f_2(1270)$, $f_1(1285)$, $a_2(1320)$, $f_0(1500)$. 
\item
The third one mostly contributes the leading source of $\bar KK$ pairs like $\phi (1020)$, $f_1(1420)$, 
$\eta (1475)$.  
\item
There is a fourth one, 
where we know little about the landscape: $a_1(1260)$, $\eta (1405)$, $a_0(1450)$, 
$\eta_2(1645)$, $f_0(1710)$, $\pi (1800)$, $f_2(1950)$. We need more data at low energies and 
probe with refined analyses. 
\end{itemize}
It might show the connection of low energy collisions of strong forces with weak dynamics. 
More data and/or analyses can change the situation {\em up} or {\em down} about FS with a pair of $\bar K$ \& $K$. 
Actually it might also give new information about the impact of gluonia  contributions to $\eta^{(\prime)}$ amplitudes mean that 
$|V_{ub}|_{\rm incl}$ is smaller than claimed; i.e., the discussion of $\eta^{(\prime)}$ wave functions \cite{GIULIA} 
enters a new stage here. 

\subsection{Another comment about diagrams}

We have some experience about the complex landscapes of diagrams vs. local operators. 
We can describe transitions of $\bar q q \to \bar s s$, $q=u,d$ with a local operator and 
used for $s \to q \bar q s$ or $q \to  s \bar s q$. However the situation is very different for suppressed 
semi-leptonic $B$ amplitudes. Diagrams can show (strong) re-scattering 
\cite{1988BOOK,WOLFFSI,CPBOOK}. However the latter 
cannot be described with local operators, while the impact of re-scattering is crucial. 

We can see other examples from non-leptonic transitions, namely to look at the data of 
$B \to2 \pi$, $3 \pi$ \& $4\pi$: BR$(B^0 \to \pi^+\pi^-) =$ $(5.12 \pm 0.19 )\cdot 10^{-6}$, 
BR$(B^0 \to \rho^0\pi^0)$ = $(2.0 \pm 0.5) \cdot 10^{-6}$,  
BR$(B^0 \to \rho^{\mp}\pi^{\pm})$ = $(2.30 \pm 0.23) \cdot 10^{-5}$, 
BR$(B^0 \to 2\pi^+ 2\pi^-)$ $<$ $1.93\cdot 10^{-5}$ and 
BR$(B^+ \to \rho^0\pi^+)$ = $(1.52 \pm 0.14) \cdot 10^{-5}$. 
It hardly suggests we can describe 
this landscape with local operators. 

\subsection{Exclusive semi-leptonic $B^0_s$ decays with $\Delta S=-1$}

So far we have hardly any information about BR$(B^0_s \to l^+ \nu X_u^{(s)}(\Delta S=-1))$  -- 
actually even for predicted rates:   
$X_u^{(s)}(\Delta S=-1) = [K^-/K_S\pi^-/K^-\pi^+\pi^-/K^-K^+K^-/K^-K_SK^+\pi^-/...]$. 
Can one find $B^0_s \to l^+\nu K^-K^+K^-$ due to re-scattering? Possible, however 
PDG2014 shows no sign for resonances in the region of 1 - 2 GeV leading to 
hadronic FS with $K \bar K K$ (except $\phi$) and $K \bar K K \pi$.  
Of course, we have to probe $B^0_s \to l^+\nu K^-\pi^+\pi^-$. It was pointed out to analyze 
$B_s^0 \to l^+ \nu K^{*-}$ and compare with $B \to K^*l^+l^-$ \cite{FELD}. 
Furthermore we have to include also broad resonances. 

The landscape of suppressed semi-leptonic $B_s^0$ decays is simpler than in $B^{0,+}$ ones. 
Therefore one can compare the numbers $|V_{ub}|_{\rm incl}$ vs. $|V_{ub}|_{\rm excl}$ 
from $B^0_s$ decays with $|V_{ub}|_{\rm incl}$ vs. $|V_{ub}|_{\rm excl}$ ones.

\subsection{Short comment about $\Lambda_b^0 \to l^- \bar \nu [p...]$}
\label{BARYON}

It is unlike that Belle II will go after beauty baryons. However, LHCb has measured 
BR$(\Lambda_b^0 \to \mu^- \bar \nu p)   = (3.9 \pm 0.8) \cdot 10^{-4}$ \cite{PATRICK}. It is the 
first semi-leptonic suppressed decays of heavy flavor baryons in general. That is quite an achievement 
itself. 

On the other hand I am not sure about the value given $|V_{ub}| = (3.27 \pm 0.15 \pm 0.17 \pm 0.06) \cdot 10^{-3}$ 
is close to real value. It depends very much what LQCD gives us quantitatively. 
It has to be tested by measuring   
$\Lambda_b^0 \to l^- \bar \nu \; p \pi^+\pi^-$, $l^- \bar \nu \; p K^+K^-$, but we should not just trust simulations.

\subsection{About the future of dispersion relations}

If transitions of $B^- \to l^- \bar \nu \; \pi^+\pi^-/K^+K^-$ and/or $\bar B^0 \to l^-\bar \nu \; \pi^+\pi^-\pi^+/K^+K^- \pi^+$ 
were found, we make a strong case to use refined dispersion relations 
to predict semi-quantitatively \cite{DR,KUBIS,KANG}. As said before, it makes much more 
work and time, but it would be us a prize, namely to change a idea into an understand not only 
non-perturbative QCD, but solve a problem about the connection of the CKM matrix for flavor dynamics. 

The authors of Ref.\cite{KANG} made reasonable assumptions. However in my view they are not truly
model-independent now. For examples: (a) They assumed that $\eta^{(\prime)}$ are describe by $\bar q_iq_i$, 
but not with two (constitute) gluon states;  
it was pointed out in Ref.\cite{GIULIA} that we need more discussions \& more data. 
(b) Can one ignore re-scattering $\pi \pi \to \bar KK$ on the very limited parts of the FS? It is an item about duality. 
It was pointed out in very recent review Ref.\cite{PhBFacts} in details about $|V_{ub}|_{\rm incl}$. Of course, Ref.\cite{KANG} 
was submitted before the review "The Physics of the B Factories"; however, 
the very active discussions have been going on about very subtle items for a long time. 

It seems to me that the meaning of `duality' is somewhat different in the world of hadrodynamics. 
For example, one can look at diagrams like $\bar ss \to f_0(980) \to \pi \pi$; one can try to describe them with "effective operators", 
but only in a small region and depend on chiral symmetry, not in general. 
Or one can discuss spectroscopies of $D^{(*)}\bar D^{(*)}$ or $B^{(*)}\bar B^{(*)}$ close to thresholds 
\cite{HANH}. 
It is not clear to me, why one can use bare poles of heavy flavor mesons, when one goes for accuracy;   
it is the opposite for the world of quarks. Furthermore the situation is much more subtle, when one 
discuss weak transitions. Or one can look at the diagrams in Fig. 2 of Ref.\cite{KANG}; 
hadrons are shown there, but not (anti-)quarks.

Even so, it shows the bridge between Hadrodynamics and HEP, but "a lot of water has still passing under the bridge"; 
i.e., it needs much more work, but also connections between Hadrodynamics and HEP.

\section{Summary}

The difference between $|V_{ub}|_{\rm excl}$ vs. $|V_{ub}|_{\rm incl}$ values seems to be sizable 
after many discussions. I suggest that real value of 
$|V_{ub}|_{\rm incl}$ might be smaller and makes the gap smaller with $|V_{ub}|_{\rm excl}$. My main points are: 
\begin{itemize}
\item 
The landscape of SM suppressed semi-leptonic $B \to l \nu X_u$ is more complex than expected before 
and more subtle due to the limits or violations of duality close to thresholds. 
\item
There is not a true prediction. This idea can be found to be incorrect, or its impact is tiny -- or it has a 
chance to put us on the right roads. If so, we cannot ignore FS with $K \bar K$, $K \bar K + \pi$'s 
in general, although their rates are smaller than with $\pi$'s. One can describe this landscape with 
somewhat different words: 
Re-scattering/FSI is important due to non-perturbative QCD.

\item 
I am not saying that these inclusive ones 
can be measured now or `soon'. However in the future LHCb and later Belle II can test this idea by probe exclusive 
one, namely $B^+ \to l^+ \nu K^+K^-$ \& $B^0 \to l^+ \nu K^+K^-\pi^-$ and maybe $B_s^0 \to l^+\nu K^-K^+K^-$.

Those enhance the averaged mass of $X_u (\Delta S=0)$, have impact of $q^2$ and 
$\Gamma (B^{0,+} \to l \nu X_u(\Delta S=0))$, $\frac{d}{dE_l}\Gamma (B^{0,+} \to l \nu X_u)$ $\leftrightarrow$ low orders of moments. We can do it also for $X_u^{(s)} (\Delta S=-1)$ 
in different landscapes. 

\item 
There are very good reasons to probe light resonances of $\Delta S=0$ with sizable impact of pairs of $\bar KK$
in the region of 1 - 2 GeV with 
more data and more refined analyses. It might narrow the gap between 
$|V_{ub}|_{\rm excl}$ vs. $|V_{ub}|_{\rm incl}$.  

\item 
Exclusive $B^{0,+} \to l \nu \, 3\,\pi/4\, \pi/\pi \eta^{(\prime)}$ 
have to be probed as much as possible, although the impact of thresholds is less. 

\item 
The future situation is simpler for $B_s^0 \to l^+\nu X_u^{(s)}(\Delta S = -1)$. I see no reason why 
$B_s^0 \to l^+\nu K^-K^+K^-$ give sizable impact on our understanding of fundamental dynamics. I think, 
we might hardly see a difference in measured values of $|V_{ub}|_{\rm incl}$ vs. $|V_{ub}|_{\rm excl}$ 
in suppressed semi-leptonic $B^0_s$ decays.  

\item
Model-insensitive analyses are the important second step, but not the final one. Experience tells us before that the `best' fitted analyses 
often do not give the best understanding of the underlying dynamics. 
The situations is probably different for $\bar bb$, $\bar cc$ and  in particular $\bar ss$ states close to thresholds. 
The landscape of "hadron-quark duality" is much more complex, namely the connection or not of local vs. 
non-local operators. Often "effective operators" are used; however often their impact are not clear (at best) 
beyond diagrams. It was  
mentioned in the `Preface' of the Memorial Book for Kolya Uraltsev \cite{MEMORIAL} and discussed in 
some contributions there.


\item 
We have refined theoretical tools (like dispersion relations) to apply to solve the "Problem' in the difference of $|V_{ub}|_{\rm incl}$ vs. $|V_{ub}|_{\rm excl}$. 
However, it takes a lot of work to do that one way or another. It would not be fair to suggest a large project about inclusive transitions, unless we have data 
from exclusive ones showing that we are on the correct `road'. Furthermore we have to measure  
$B^+ \to l^+\nu K^+K^-$ and $B^0 \to l^+\nu K^+K^-\pi^-$; likewise for $B^0_s \to l^+\nu K^-$
\& $B^0_s \to l^+\nu K^-K^+K^-$ and compare the results.

\end{itemize} 
At least we get novel lessons about non-perturbative QCD -- and possibly beyond. 
These suggestions fall first on the shoulders of my experimental 
colleagues. It is not unusual how theorists work.

\vspace{2mm}

{\bf Acknowledgments:} This work was supported by the NSF under grant numbers PHY-1215979 \& PHY-1520966. 
I am grateful to the Mainz Institute for Theoretical Physics ({\bf MITP}) and the Bonn 
Universit$\ddot {\rm a}$t for their hospitality and 
their partial support.

\vspace{2mm}



\begin{thebibliography}{99}

\bibitem{PhBFacts} 
Adrian Bevan {\em et al.}, "The Physics of the B Factories", {\em Eur.Phys.J.} {\bf C74} (2014) 11, 3026; 
arXiv:1406.6311 [hep-ex], Sects. 17.1.1 - 17.1.6., pps. 186 -- 215. 



\bibitem{MANNEL} 
S. Faller {\em et al.}, {\em Phys.Rev.} {\bf D89} (2014) 014015; arXiv:1310.6660 [hep-ph]. 

\bibitem{MEMORIAL}
"QCD AND HEAVY QUARKS -- In Memorial Nikolai Uraltsev"; I.I. Bigi, P. Gambino, Th. Mannel (editors), 
World Scientific, 2015. 

\bibitem{KANG}
X.-W. Kang {\em et al.}, {\em Phys.Rev.} {\bf D89} (2014) 053015; arXiv:1312.1193 [hep-ph].

\bibitem{STEFAN}
A. Crivellin, St. Pokorski, {\em Phys.Rev.Lett.}{\bf 114} (2015) 1, 011802, arXiv:1407.1320 [hep-ph].  

\bibitem{74}
I.I. Bigi, N. Uraltsev, {\em Nucl.Phys.} {\bf B423} (1994) 33; {\em Z.Phys.} {\bf C62} (1994) 623.

\bibitem{94} 
D. Benson, I. Bigi, N. Uraltsev, {\em Nucl.Phys.} {\bf B710} (2005) 371; 
I. Bigi, N. Uraltsev, {\em Phys.Lett.} {\bf B579} (2004) 340. 

\bibitem{PAOLO}
P. Gambino, G. Ossola, N. Uraltsev, {\em JHEP} {\bf 0509} (2005) 010; 
P. Gambino {\em et al.}, {\em JHEP} {\bf 0710} (2007) 058. 


\bibitem{MISHA}
M.A. Shifman, A.I. Vainshtein, V.I. Zakharov, {\em Nucl. Phys.} {\bf B147} (1979) 385;
V.A.Novikov, M.A.Shifman, A.I.Vainshtein and V.I.Zakharov, Nucl.Phys. B249 445 (1985);  
B. Chibisov, R. Dikeman, M. Shifman, N.G. Uraltsev, {\em Int. Journ. Mod. Phys.} {\bf A12} (1997) 2075; 
M. Shifman, in: {\em Boris Ioffe Festschrift “At the Frontier of Particle Physics -- Handbook of QCD”}, Ed. M. Shifman (World Scientific, Singapore, 2001), Vol. 3, p. 1447; hep-ph/0009131.

\bibitem{TWOROADS}
I. Bigi, Th. Mannel, N. Uraltsev, {\em JHEP} {\bf 1109} (2011) 012; 
I. Bigi, Th. Mannel, S. Turczyk, N. Uraltsev , {\em JHEP} {\bf 1004} (2010) 073.


\bibitem{VADEMECUM}
I.I. Bigi, N. Uraltsev, {\em Int.J.Mod.Phys.}{\bf A16} (2001) 5201-5248; 
I.I. Bigi, Th. Mannel, arXiv:hep-ph/0212021.

\bibitem{POLEQUARK}
M.A. Shifman, contribution to "QCD AND HEAVY QUARKS -- In Memorial Nikolai Uraltsev", 
World Scientific, 2015, arXiv:1310.1966; 
I.I. Bigi, M.A. Shifman, N.G. Uraltsev, {\em Ann.Rev.Nucl.Part.Sci.} {\bf 47} (1997) 591-661, 
hep-ph/9703290; 
I.I. Bigi, M.A. Shifman, N.G. Uraltsev, A.I. Vainshtein, {\em Phys.Rev.} {\bf D50} (1994) 2234-2246, 
 hep-ph/9402360. 


\bibitem{DR}
M.R. Pennington, Proceed. of MESON2002, arXiv:hep-ph/0207220;
J.R. Pelaez {\em et al.}, Proceed. of the Hadron 2011 [arXiv:1109.2392];
J.R. Pelaez, arXiv:1301.4431.

\bibitem{1988BOOK}
I.I. Bigi, V.A. Khoze, N.G. Uraltsev, A.I. Sanda, p. 175-248 in "CP Violation", C. Jarlskog (Editor), 
World Scientific (1988).


\bibitem{WOLFFSI}
L. Wolfenstein, {\em Phys.Rev.} {\bf D43} (1991) 151.

\bibitem{CPBOOK} 
I.I. Bigi, A.I. Sanda, "CP Violation, Second Edition", 
{\em Cambridge Monographs on Particle Physics, Nuclear Physics and Cosmology} 
(Cambridge University Press) 2009, pp. 62 - 69, Sect. "4.10 The role of final interactions". 


\bibitem{GIULIA}
C. Di Donato, G. Ricciardi, I. Bigi, {\em Phys.Rev.} {\bf D85} (2012) 013016, 
arXiv:1105.3557 [hep-ph]. 




\bibitem{FELD} 
Th. Feldmann, B. Mueller, D. van Dyk, {\em Phys.Rev.} {\bf D92} (2015) 3, 034013;  
arXiv:1503.090630 [hep-ph].



\bibitem{PATRICK}
LHCb collab., R. Aaij {\em et al.}, arXiv:1504.01568 [hep-ex]. 




\bibitem{KUBIS}
Chr. Hanhart,  {\em Proceedings of CHARM 2013},
"Modelling low-mass resonances in multi-body decays" [arXiv:1311.6627];
S. Gardner, U.-G. Meissner, {\em Phys.Rev.} {\bf D65} (2002) 094004 [arXiv:hep-ph/0112281]. 


\bibitem{HANH}
C. Hanhart {\em et al.}, arXiv:1507.00382v1 [hep-ph]. 




\end{thebibliography}
\end{document}